\documentclass[twocolumn, amsmath,amssymb,superscriptaddress,aps,prl]{revtex4-2}
\usepackage{graphicx}
\usepackage{epstopdf}
\usepackage{dcolumn}
\usepackage{bm}
\usepackage{subfigure}
\usepackage{braket}
 
\usepackage{color}
\usepackage{hyperref}
\usepackage[nameinlink,capitalise]{cleveref}
\begin{document}
\begin{titlepage}
\title{Laser-\color{black}Controlled\color{black}\,
 Nonlinear Hall Effect in Tellurium Solids \\via Nonlinear Phononics}
\author{Hongyu Chen} 
\affiliation{Shenzhen Geim Graphene Center and Institute of Materials Research, Tsinghua Shenzhen International Graduate School, Tsinghua University, Shenzhen 518055, People's Republic of China}
\author{Xi Wu} 
\affiliation{State Key Laboratory of Low Dimensional Quantum Physics and Department of Physics, Tsinghua University, Beijing 100084, People's Republic of China}
\author{Jiali Yang} 
\affiliation{Shenzhen Geim Graphene Center and Institute of Materials Research, Tsinghua Shenzhen International Graduate School, Tsinghua University, Shenzhen 518055, People's Republic of China}
\author{Peizhe Tang}
\email{peizhet@buaa.edu.cn}
\affiliation{School of Materials Science and Engineering, Beihang University, Beijing 100191, People's Republic of China}
\affiliation{Institute for Advanced Study, Tsinghua University, Beijing 100084, People's Republic of China}
\affiliation{Frontier Science Center for Quantum Information, Beijing 100084, People's Republic of China}
\author{Jia Li}
\email{li.jia@sz.tsinghua.edu.cn}
\affiliation{Shenzhen Geim Graphene Center and Institute of Materials Research, Tsinghua Shenzhen International Graduate School, Tsinghua University, Shenzhen 518055, People's Republic of China}
\date{\today}
\begin{abstract}


\color{black}A Terahertz (THz) laser with strong strength could excite more than one phonons and induce a transient lattice distortion termed as nonlinear phononics.\color{black}\, This process allows dynamic control of various physical properties, including topological properties. Here, using first-principles calculations and dynamical simulations, we demonstrate that THz laser excitation can modulate the electronic structure and \color{black}the signal of nonlinear Hall effect\color{black}\, in elemental solid tellurium (Te). By strongly exciting the chiral phonon mode, we observe a non-equilibrium steady state characterized by lattice distortion along the breathing vibrational mode. This leads to a transition of Te from a direct to an indirect semiconductor. In addition, the energy dispersion around the Weyl point is deformed, leading to variations in the local Berry curvature dipole. As a result, the nonlinear Hall-like current in Te can be modulated with electron doping \color{black} where the sign of current could be reversed under a strong THz laser field.\color{black}\, Our results may stimulate further research on coupled quasiparticles in solids and the manipulation of their topological transport properties using THz lasers.
\end{abstract}
\maketitle
\end{titlepage}

\newpage

\color{black} The atomic displacements within the lattice distortion induced by nonlinear phononics are corresponding to the lattice vibration along the Raman mode \cite{Forn2019,Rivera2020,Liu2010,Forst2011}. These THz laser-driving atomic displacements can survive\color{black}\, within the phonon lifetime, resulting in transient lattice distortions\cite{Neufeld2022,Caruso2023,Fechner2024,Subedi2014,Mankowsky2015,Fechner2016,Juraschek2017,Frenzel2023}. Hence, \color{black} nonlinear phononics could\color{black}\, enables dynamic structure design and facilitates ultrafast \color{black}phase transitions of electronic structures such as Weyl point creation or annihilation\cite{Sie2019}, transient magnetic order\cite{Nova2017,Disa2020,Afanasiev2021}, superconducting order\cite{Tang2023,Mankowsky2014} and spin-orbital order\cite{Radhakrishnan2024}. In stead of phase transitions, nonlinear phononics can switch crystals into some non-equilibrium steady states which are rarely detected in the original ground states such as ideal Weyl semimetals\cite{Shin2024,Ma2017}, excited magnons\cite{Mashkovich2021}, ferroelectricity\cite{Fechner2024} and its polarization reversion\cite{Subedi2015,Mankowsky2017}.\color{black}\,

\color{black}Recently, the magneto-optic Kerr effect is reported to be able to be controlled by THz laser\cite{Frenzel2023}, emphasizing the importance of studying \color{black}crystals' responses to multiple physical fields\color{black}\,. As a nonlinear response, the nonlinear Hall effect (NHE) can reflect the topological properties of the Fermi surface and depends on the product of the Berry curvature and the Fermi velocity, known as the Berry curvature dipole (BCD)\cite{Sodemann2015,Wang2023}. Unlike the quantum spin Hall effect\cite{Kato2004} and the quantum anomalous Hall effect\cite{Chang2013}, the NHE is characterized by a transverse Hall voltage in response to two longitudinal currents, requiring the breaking of inversion symmetry rather than time-reversal symmetry\cite{Woltgens1993,Yang2008,Zhang2009,Dugaev2012,Sodemann2015}. On the other hand, simple elemental solids such as arsenic (As)\cite{Hossain2024}, bismuth (Bi)\cite{Drozdov2014,Schindler2018}, and tellurium (Te)\cite{Tsirkin2018,Gatti2020,Sakano2020} exhibit a variety of topological properties, making them good platforms for exploring the topological response under external fields. Among these materials, elemental Te has been extensively studied in the context of Weyl semimetals\cite{Gatti2020,Sakano2020}, current-induced magnetism\cite{Furukawa2021,Shalygin2012}, gyrotropic effect and \color{black}NHE\cite{Tsirkin2018}, and topological phase transitions induced by external fields\cite{Agapito2013,Hirayama2015,Ning2022,Jnawali2020}. Therefore, Te is an excellent candidate for the study of controlling NHE under THz laser excitation.\color{black}\,

In this Letter, we investigate \color{black}Te's response to THz lasers and the corresponding NHE \color{black}\,using first-principles calculations and dynamical simulations. We found that the chiral infrared-active mode of Te can be directly excited by the THz laser pump, which is coupled to an indirectly excited Raman-active mode. \color{black}Through numerical and analytical demonstrations of the dynamical regimes for the two coupled phonons, we find that a lattice distortion along this Raman mode\color{black}\,without symmetry breaking can be transiently stabilized within the phonon lifetime. This lattice distortion induces a deformation of the \color{black} conduction-band edge of Te, which leads \color{black}\,to a transition from a direct to an indirect semiconductor. Furthermore, \color{black} the sign of NHE signal in Te with the electron doping can be reversed under a strong laser field\color{black}\,, resulting from the varying distribution of the BCD induced by the lattice distortion. Our results pave the way for \color{black} chiral solids' multiple responses to external fields including ultrafast dynamical lattice control under THz laser and the nonlinear electrical transport.\color{black}\,

\begin{figure}[pt]
	\includegraphics[scale=1]{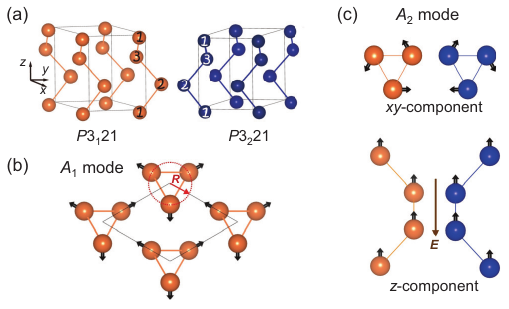}	
	\caption{\label{1} Crystal structures and phonon modes of Te. (a) Lattice configurations of right-handed (orange spheres) and left-handed (blue spheres) elemental Te, with each unit cell containing three atoms, labelled with black and white numbers. (b) Raman-active phonon mode $A_1$, where the radius of the spiral chain is denoted by $R$. (c) Infrared-active phonon mode $A_2$ in right-handed and left-handed elemental Te. The top panel shows the $xy$ component and the bottom panel the $z$ component, which is responsible for the dipole coupling with the electric field of the THz laser.}
\end{figure}

The elemental solid Te has two enantiomorphic crystal structures \color{black}with the space groups of $P3_121$ and $P3_221$ and point group of $D_3$. As shown in Fig. 1(a), the helical atomic chain of Te with three-fold screw symmetry along the $z$-direction indicates \color{black}the inversion symmetry breaking and chiral nature\color{black}\,. Among the six vibrational phonon modes of Te (Fig. S1 and Tab. S1 \cite{mine}), the $A_1$ mode is nonpolar with an in-plane breathing lattice vibration in the $xy$ plane [Fig. 1(b)], and the $A_2$ mode is chiral with an in-plane rocking lattice vibration [Fig. 1(c)]. However, the $z$-component of the $A_2$ mode is polar and can couple with the electric field of the pumped THz laser. \color{black}Since the irreducible representations in $D_3$ point group obey $A_2\otimes A_2\otimes A_1\otimes, ..., \otimes A_1 \supset A_1$, \color{black}\,once the infrared-active $A_2$ mode is excited by laser pumping, the coupled $A_1$ mode can be indirectly excited without breaking the overall symmetry $A_1$. \color{black} Then we select infrared-active $A_2$ mode and Raman-active $A_1$ mode as the two coupled phonons.\color{black}


The dynamic regimes of the $A_1$ and $A_2$ modes were first evaluated individually. As shown in Fig. 2(a), the potential curve of the $A_2$ mode is parabolic and symmetrical. In contrast, the $A_1$ mode has an asymmetric potential curve with respect to positive and negative vibrations, as shown in Fig. 2(b). The potential of the $A_1$ mode can be expressed as:
\begin{equation}
	\begin{aligned}
		V(Q_{A_1})=\frac{1}{2}\omega_{A_1}^2Q_{A_1}^2 +\frac{1}{3}a_3Q_{A_1}^3+\frac{1}{4}a_4Q_{A_1}^4+...\frac{1}{n}a_nQ_{A_1}^n...
	\end{aligned}
\end{equation}
where $Q_{A_1}$ is the normal coordinate of the $A_1$ mode, and $\omega_{A_1}^2$ and $a_n$ are the parameters related to the harmonic and anharmonic terms. The values of $\omega_{A_1}^2$ and $a_n$ can be obtained by fitting the potential curve $V(Q_{A_1})$. As shown in Tab. 1, the absolute value of $a_3$ is 7.62 (meV/$\rm amu^{\frac{3}{2}}$\AA$^3$), which is significant compared to $\omega_{A_1}^2$ 38.74 (meV/amu\AA$^2$) and much larger than $a_4$ 0.24 (meV/$\rm amu^{2}$\AA$^4$), \color{black}indicating that the intrinsic vibration of $A_1$ mode is not simple harmonic and the analytical solution possesses constant term and higher-order harmonics (see in Sec. II of the SI \cite{mine})  which can be written as 
\begin{equation}
	\begin{aligned}
		Q_{A_1}^{(2)}=-\frac{a_3A^2}{2\omega_{A_1}^2}+\frac{a_3A^2}{6\omega_{A_1}^2}\cos{2\omega{t}}
	\end{aligned}
\end{equation}
where $A$ and $\omega$ refer to the amplitude and renormalized frequency of the $A_1$ mode, respectively. The constant in the solution from the significant term $a_3Q_{A_1}^3$ \color{black}\,implies an absolute shift in the lattice equilibrium position once the $A_1$ mode is excited. Since $a_3$ $\textless$ 0, the shift can be viewed as an expansion of the radius $R$ of the helical chain, as shown in Fig. 1(b).




\begin{table}[pt]
	\centering
	\caption{\label{t1} Coefficients in the total potential energy $V(Q_{A_2},Q_{A_1})$ that fitted from first-principle calculations.}
		\begin{ruledtabular}
		\begin{tabular}{p{4cm}<{\centering}p{4cm}<{\centering}}
			$\omega_{A_2}^2$ (meV/amu\AA$^2$) &26.72\\
			$b_4$ (meV/$\rm amu^{2}$\AA$^4$)&0.28\\
			$\omega_{A_1}^2$ (meV/amu\AA$^2$)&38.74\\
			$a_3$ (meV/$\rm amu^{\frac{3}{2}}$\AA$^3$) &-7.62\\
			$a_4$ (meV/$\rm amu^{2}$\AA$^4$)&0.24\\
			$g_1$ (meV/$\rm amu^{\frac{3}{2}}$\AA$^3$)&-0.85\\
			$g_2$ (meV/$\rm amu^{2}$\AA$^4$)&-2.02\\
		\end{tabular}
		\end{ruledtabular}
\end{table}

\begin{figure}[pt]
	\includegraphics[scale=1]{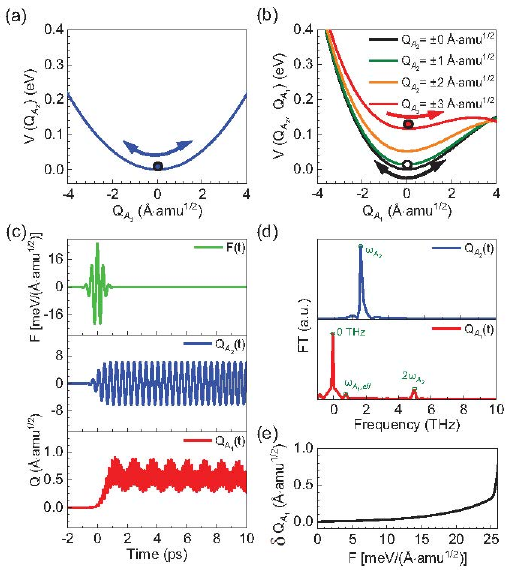}	
	\caption{\label{2} Dynamical regime of the THz laser-driven coupling between the $A_1$ and $A_2$ modes. (a) Potential energy curves for the lattice vibrations of the $A_2$ mode. (b) Potential energy curves for the lattice vibrations of the $A_1$ mode under varying amplitudes of the lattice vibrations induced by the $A_2$ mode. (c) Simulated time-dependent amplitudes of the laser pump (F), lattice vibrations of the $A_2$ mode ($Q_{A_2}(t)$) and lattice vibrations of the $A_1$ mode ($Q_{A_1}(t)$). (d) Fourier transform of $Q_{A_2}(t)$ and $Q_{A_1}(t)$. \color{black}(e) Shift of the equilibrium positions of lattice vibrations along $A_1$ mode as a function of the pump strength.\color{black}}
\end{figure}

\begin{figure}[pt]	\includegraphics[scale=1]{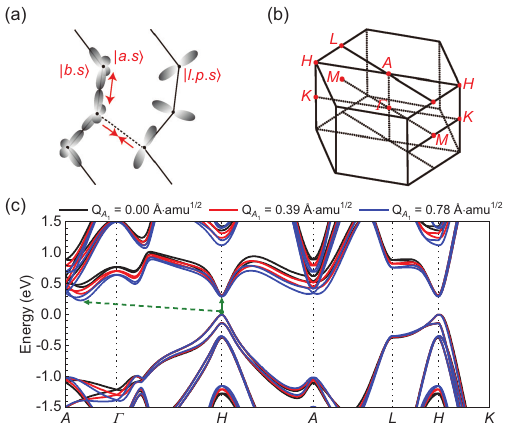}
	\caption{\label{3} THz laser-induced changes in the chemical bonding and electronic structure of chiral elemental Te. (a) Illustration of the $p$ orbital wavefunctions for three types of states around the Fermi level: bonding, antibonding and lone-pair states, labelled $\ket{b.s}$, $\ket{a.s}$ and $\ket{l.p.s}$, respectively. (b) Brillouin zone of elemental Te. (c) Band structures of elemental Te under different amplitudes of lattice vibrations induced by the $A_1$ mode. The values of $Q_{A_1}$ equal to 0.39 and 0.78 \AA$\sqrt{amu}$ correspond to shifts in the equilibrium position of Te of 0.1 and 0.2 \AA, respectively.}
\end{figure}

Before discussing the laser-induced dynamical behavior of Te, we describe the nonlinear phonon interactions within Te. These interactions can be understood from the potential curve of the $A_1$ mode \color{black}as shown in Fig. 2(b). As the displacements of $Q_{A_2}$ increase, the asymmetry of $A_1$'s potential curves becomes more pronounced. These results suggest that resonant pumping of the $A_2$ mode can significantly alter the potential curve of the $A_1$ mode. Then we write the total potential of the coupled $A_1$ mode and $A_2$ mode as: 
\color{black}\,
\begin{equation}\label{totalpotential}
	\begin{aligned}
		V(Q_{A_2}, Q_{A_1}) &= \frac{1}{2}\omega_{A_2}^2Q_{A_2}^2 + \frac{1}{4}b_4Q_{A_2}^4\\
		&+ \frac{1}{2}\omega_{A_1}^2Q_{A_1}^2 + \frac{1}{3}a_3Q_{A_1}^3 + \frac{1}{4}a_4Q_{A_1}^4\\
		&+ g_1Q_{A_2}^2Q_{A_1} + \frac{1}{2}g_2Q_{A_2}^2Q_{A_1}^2\\
	\end{aligned}
\end{equation}
Here, $Q_{A_1}$ and $Q_{A_2}$ are the normal coordinates of the $A_1$ and $A_2$ modes, respectively. $\omega_{A_1}^2$ and $\omega_{A_2}^2$ are associated with the harmonic terms of the $A_1$ and $A_2$ modes, respectively. $b_4$ corresponds to the anharmonic term of the $A_2$ mode, while $a_3$ and $a_4$ correspond to the anharmonic terms of the $A_1$ mode. $g_1$ and $g_2$ represent the symmetry-allowed coupling terms between the $A_1$ and $A_2$ modes. All the parameters were determined by fitting the potential curve, as shown in Tab. 1. 

\begin{figure*}[t]
	\includegraphics[scale=1]{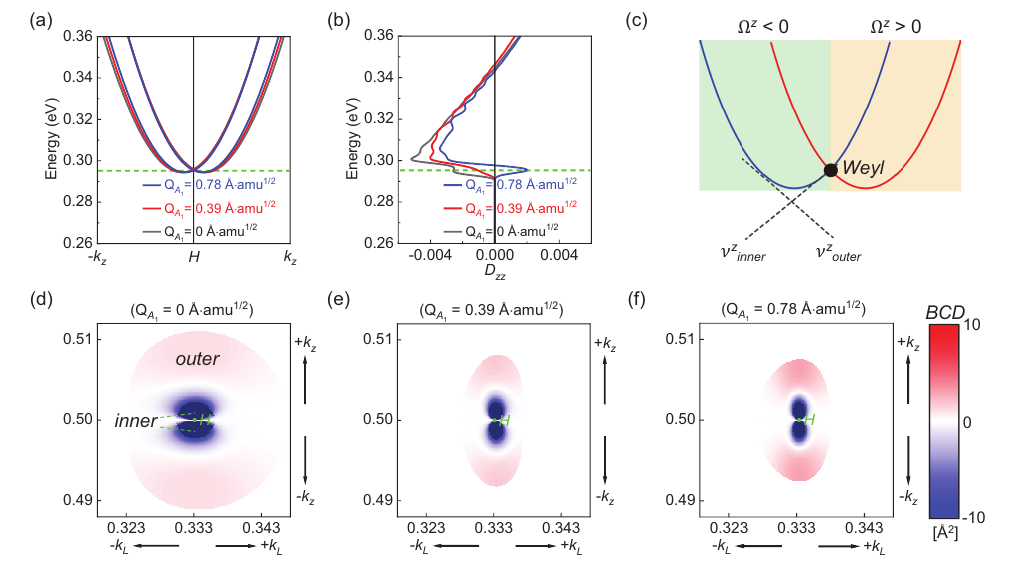}
	\caption{\label{4} THz laser-controlled nonlinear Hall effect around the Weyl point in electron-doped elemental Te. (a) Band structures around the Weyl point at the $\bm{H}$ point. (b) $D_{zz}$ component of the BCD tensor in elemental Te under different amplitudes of lattice vibration induced by the $A_1$ mode. (c) Schematic illustration of the signs of the Berry curvature and Fermi velocity around the Weyl point at the $\bm{H}$ point. The Berry curvature and Fermi velocity have opposite signs in the $+k_z$ and $-k_z$ regions, with the Fermi velocities in each region having opposite signs near the $\bm{H}$ point ($v^{z}_{inner}$) and farther from the $\bm{H}$ point ($v^{z}_{outer}$). (d)-(f) BCD distributions around the Weyl point in momentum space at the energy level indicated by the green dashed line in (a) for elemental Te under lattice vibration amplitudes induced by the $A_1$ mode of 0, 0.39 and 0.78 \AA$\sqrt{amu}$, respectively. The values of $Q_{A_1}$ equal to 0.39 and 0.78 \AA$\sqrt{amu}$ correspond to shifts in the equilibrium position of Te of 0.1 and 0.2 \AA, respectively.}
\end{figure*}

\color{black}
Hence, the ultrafast dynamic lattice control of Te induced by a THz laser can be described by solving the coupled equations of motion from the canonical transformation of Eq. \ref{totalpotential}. The external driving force from the oscillating electric field is applied solely in the equation motion of $A_2$ mode (shown in Sec. III of SI\cite{mine}). As shown in the up and middle panels of Fig. 2(c), the infrared-active $A_2$ mode is excited when the resonant laser pumping is applied. The Fourier transformation of $Q_{A_2}(t)$ is shown in the up panel of Fig. 2(d), where the only peak is corresponding to the applied frequency, verifying the resonant excitation of $A_2$ mode. Then, in $A_1$ mode's equation of motion, the nonzero $Q_{A2}^2$ term makes the initial solution for $Q_{A_1}(t)$ nonzero which lets $A_1$ mode also start vibrating even without an external driving force as shown in the bottom panel of Fig. 2(c) where we can see the unconventional beat and equilibrium-position shift. The Fourier transformation of $Q_{A_1}(t)$ is shown in the bottom panel of Fig. 2(d) where three frequencies emerge: zero frequency which is corresponding to the equilibrium-position shift, a frequency equaling to double $\omega_{A_2}$ from $g_1Q_{A_2}^2Q_{A_1}$ term denoted as $2\omega_{A_2}$ and a approaching-zero frequency related to $\omega_{A_1}$ denoted as $\omega_{A_1,\rm eff}$ (the analytical derivation process is shown in Sec. III of SI\cite{mine}). The pump strength-dependence shift of equilibrium-position of $\delta Q_{A_1}$ is shown in Fig. 2(e) where $\delta Q_{A_1}$ increases monotonically and indicates the lattice distortion along $A_1$ mode. Strikingly, the lattice distortion discovered here is consistent with that in $A_1$ mode's intrinsic vibration. Hence, the anharmonic nature of $A_1$ mode and the coupling terms with $A_2$ mode are both the reasons of the shift of equilibrium position in $A_1$ mode's vibration. Considering the unit conversion with Te atom's mass, Born effective charge and experimentally achievable pump strength\cite{Peng2015,Shalaby2015}, the feasible shift of the equilibrium position of Te atoms induced by the vibration of the $A_1$ mode is about 0.8 \AA$\sqrt{amu}$.
\color{black}\,

After studying the origin of the THz laser-induced lattice distortion in elemental Te, the effect of this distortion on its electronic structure was investigated. \color{black}It is revealed that the lattice distortion in elemental Te induced by expansion of the radius $R$ of the helical chain [see Fig. 1(b)] is in the structural phase located at the path from $P3_121$ to $R\overline{3}m$ without symmetry breaking, as shown in Fig. S2 in SI \cite{mine}. As shown in Fig. 3(a), the electronic states around the Fermi level, derived from the $p$ orbitals of Te, consist of three groups with respect to the classification of atomic chemical bonding:\color{black}\, the intrachain bonding state, the intrachain antibonding state (originating from the $p_{x/y}$ orbitals), and the interchain lone-pair state (originating from the $p_z$ orbitals), denoted as $\ket{b. s.}$, $\ket{a.s.}$, and $\ket{l.p.s.}$, respectively [\cite{Hirayama2015}]. In the initial undistorted structure of Te, the valence bands mainly originate from $\ket{l.p.s.}$, while the conduction bands mainly originate from $\ket{a.s.}$, with both the valence band maximum (VBM) and the conduction band minimum (CBM) located at the $\bm{H}$ point [Figs. 3(b) and 3(c)]. However, under the unidirectional lattice expansion of Te as shown in Fig. 1(b), the intrachain atomic spacing increases, while the interchain atomic spacing decreases [Fig. 3(a)]. \color{black} We also observe\color{black}\,a decrease in the energy of the conduction band edge along the $\bm{\Gamma}$-$\bm{A}$ direction, resulting in a lower energy state compared to the conduction band at the $\bm{H}$ point. Consequently, this transition shifts the elemental Te from a direct to an indirect semiconductor, as shown in Fig. 3(c). 


\color{black}
Finally, the NHE response of elemental Te under the influence of a THz laser is investigated. It has been reported that the giant Berry curvature and BCD around a Weyl point at the $\bm{H}$ point in conductance band can contribute NHE current, magnitude of which is possibly to be detected in experiment \cite{Tsirkin2018}. Hence, we investigate the controlling of NHE signals of Te with the chemical potentials where the Weyl point emerges under the lattice distortions as shown in Fig. 1(b). To elevate the Fermi level to the Weyl point at the $\bm{H}$ point, electron doping is required in Te.\color{black}\, In this scenario, increasing the amplitude of the lattice vibrations induced by the $A_1$ mode could cause observable deformations in the energy dispersion of the bands around the Weyl point, as shown in Fig. 4(a). The NHE signal can be calculated by integrating the Berry curvature dipole (BCD) over the Fermi surface $S$ in momentum space $\bf{k}$ at a given Fermi energy $\mit\epsilon$. The BCD \color{black}which determines the in-plane NHE current flowing along the trigonal axis\color{black}\,[\cite{Tsirkin2018}] is given by
\begin{equation}    
	\begin{aligned}
		D_{zz}(\epsilon)=\frac{1}{(2\pi)^3}\sum_{n}\int_{E_{\bf{k}n}=\epsilon}dS{v}^{z}_{\bf{k}n}\Omega^{z}_{\bf{k}n}
	\end{aligned}
\end{equation}
where ${v}^{z}_{\bf{k}n}$ and $\Omega^{z}_{\bf{k}n}$ are the Fermi velocity and Berry curvature for the $n^{th}$ band, respectively.  As shown in Fig. 4(b), increasing the amplitude of the lattice vibrations induced by the $A_1$ mode leads to a reversal of the $D_{zz}$ component of the BCD tensor, suggesting that a reversal of the NHE could be observed in elemental Te. The reversal of the NHE signal can be understood by considering the sign of the critical Berry curvature and the Fermi velocity around the Weyl point. As shown in Fig. 4(c), there are two signs of the Fermi velocity (labelled $v^z_{inner}$ and $v^z_{outer}$) and two signs of the Berry curvature (labelled $\Omega^{z}<0$ and $\Omega^{z}>0$) on either side of the Weyl point in elementary Te. Therefore, the change in BCD can be understood as a result of the changing contributions from the inner and outer regions in the momentum space near the $\bm{H}$ point of elementary Te. As shown in Figs. 4(d) to 4(f), as the amplitude of the lattice vibration induced by the $A_1$ mode increases, the integral over the inner region decreases while the integral over the outer region remains almost unchanged, leading to a reversal of the BCD and consequently of the NHE signal as the THz laser intensity increases. This indicates that the reversal of the NHE signal in elemental Te under a THz laser is due to the rearrangement of the BCD density around the Weyl point.
\color{black}\,

In conclusion, the topological response in the chiral crystal of elemental Te can be effectively manipulated using a THz laser. In elemental Te, the harmonic infrared active phonon mode $A_2$ can be directly excited by the THz laser, which in turn indirectly excites the anharmonic Raman active phonon mode $A_1$ due to the coupling between the $A_1$ and $A_2$ modes. \color{black}The anharmonic nature of the $A_1$ mode and the coupling with $A_2$ mode\color{black}\, induces a shift in the equilibrium position during lattice vibrations, leading to lattice distortion in elemental Te. This lattice distortion reconfigures the chemical bonds and changes the electronic structure, resulting in a \color{black} transient transition of electronic structure\color{black}\, of elemental Te from a direct to an indirect semiconductor. In addition, the rearrangement of the Berry curvature dipole (BCD) density around the Weyl point, causes a reversal of the NHE signal in elemental Te. This study of the topological responses of elemental Te under THz laser excitation provides a blueprint for the laser-driven manipulation of coupled quasiparticles in elemental solids and provides insight into the control of other topology-related transport properties using THz lasers.

\section{Acknowledgments}
We thank Xiaobin Chen and Haowei Chen for fruitful discussions. This work was supported by the National Key R\&D Program of China (2021YFA1400100, 2022YFA1203400), the National Science Foundation of China (12274254), the Local Innovative and Research Teams Project of Guangdong Pearl River Talents Program (2017BT01N111), the Basic Research Project of Shenzhen, China (JCYJ20200109142816479, WDZC20200819115243002). P.T. was supported by the National Natural Science Foundation of China (no. 12234011 and no. 12374053). Computational resources were provided by the High Performance Computing Platform of Nanjing University of Aeronautics and Astronautics.

H. C. and X. W. contribute equally to this work.

$Note$ $added.$---Recently, we became aware of an e-print
with similar outcome aiming at deforming bandedge of elemental Tellurium by exciting coherent phonons ($A_{1}$ and $E^{\prime}$) \cite{gatti2024}.

\bibliography{laser}
\end{document}